\documentclass[preprint, prb, amsmath, aps, subeqn, showpacs]{revtex4}
\usepackage{graphicx}

\begin{document}

\title{Wave excitations of drifting two-dimensional electron gas
under strong inelastic scattering}

\author{V. V. Korotyeyev, V. A. Kochelap}
\affiliation{Department of Theoretical Physics,
Institute for Semiconductor Physics, Pr. Nauki 41, Kiev 03028,
Ukraine}
\author{L. Varani}
\affiliation{Institut d'\'Electronique du Sud, CNRS UMR 5214,
University Montpellier 2 France}
 will be published in Journal of Applied Physics
\begin{abstract}
We have analyzed low-temperature behavior of two-dimensional electron
gas  in polar heterostructures subjected to a high electric field.
When the optical phonon emission is the fastest relaxation process, we have
found existence of  collective wave-like excitations of the electrons.
These wave-like excitations are periodic in time oscillations of the electrons
in both real and momentum spaces. The excitation spectra are of multi-branch
character with considerable spatial dispersion. There are one acoustic-type
and a number of optical-type branches of the  spectra.  Their small damping is
caused by  quasi-elastic scattering of the electrons and formation of
relevant space charge.  Also there exist waves with zero frequency and finite
spatial periods - the standing waves. The found excitations of the electron
gas can be interpreted as synchronous in time and real space manifestation
of well-known optical-phonon-transient-time-resonance. Estimates of
parameters of the excitations for two polar heterostructures,
GaN/AlGaN and ZnO/MgZnO, have shown that excitation frequencies are in
THz-frequency range, while standing wave periods are in sub-micrometer
region.
\end{abstract}

\pacs{72.30.+q, 72.20.Ht, 73.63.Hs}
\keywords{AAA}

\maketitle
\section{Introduction}\label{Introduction}

In semiconductor materials and heterostructures at low temperature,
when $e^{-\hbar \omega_0/k_B T} \ll 1$ ($\omega_0$,  $k_B$ and $T$ are the
optical phonon frequency,  the Boltzmann constant and the
temperature, respectively) the absorption of optical phonons
by electrons is practically absent. While optical phonon emission
can be the dominant scattering mechanism for hot electrons if
electron-optical phonon coupling is strong enough.
In samples with high electron mobility, where
the electrons undergo only weak quasi-elastic scattering at
a low field, the dynamics of an electron subjected
to a steady state high electric field $F_0$ is the following.
The electron is almost ballistically accelerated by the field
until reaching the optical phonon energy. Then, the optical
phonon emission occurs so that the electron loses practically
all its energy and stops. The processes are repeated in time.
This dynamics gives rise to temporal and spatial modulation
of the electron momentum, ${\bf p}$, and velocity ${\bf v}$
with characteristic time period,
$\tau_F = p_0/e F_0$, and length, $l_F = e F_0
 \tau^2_F/2\,m^* \equiv \hbar \omega_0/e F_0$, where
 $p_0 =\sqrt{2 m^* \hbar \omega_0}$, $e$ is the elementary charge
 and $m^*$ is the electron effective mass.

For this kind of cyclic motion, the electron momentum space
can be divided into two regions. The first is the
{\em passive} region, where the electron energy
$E ({\bf p}) < \hbar \omega_0$. In this region inelastic scattering
by optical phonons is almost absent, so that electron scattering
is practically elastic and can be characterized by an elastic
scattering time, $\tau_p$.
The second region is the {\em active} region, $E ({\bf p}) > \hbar \omega_0$,
where optical phonon emission is the dominant process, i.e.
\begin{equation} \label{necessary-cond}
\tau_{op}^{em} \ll \tau_p\,,
\end{equation}
with $\tau_{op}^{em} $ being the optical phonon emission time.
Pronounced cyclic motion is possible, when
\begin{equation} \label{F-conditions}
\tau_{op}^{em} \ll \tau_F \leq \tau_p\,,
\end{equation}
i.e., the time of flight of the electrons through the passive region is shorter than
the scattering time, while penetration of the electrons into
the active region, $p - p_0 = \Delta p$,  is sufficiently small,
\begin{equation} \label{penetration}
\Delta p \sim p_0 \frac{\tau_{op}^{em}}{\tau_F} \ll p_0\,.
\end{equation}
Actually, the latter inequalities define a range of the electric fields,
where this kind of periodic motion can occur.
The critical electric field necessary for the onset of the cyclic electron
motion is estimated to be
\begin{equation} \label{cr-field}
F_{cr} = \frac{p_0}{e \tau_p}\,.
\end{equation}

The presented scenario is essentially single electron physical picture valid at
low or modest electron concentrations, when {\em e-e} collisions
do not destroy the cyclic motion.

The possibility of cyclic  acceleration-stop electron
dynamics due to strong inelastic scattering by optical phonons
was predicted six decade ago by Shockley.~\cite{Shockley}
Experimental evidences of the {\it cyclic dynamics
in real space} were found by analyzing low temperature I-V
characteristics of short diodes made from different polar materials:
InSb,\cite{InSb} InGaAs,\cite{InGaAs}  GaAs,\cite{GaAs} and
InP.\cite{InP} Tens of cycles were identified at low
temperatures. Very recently, comprehensive Monte-Carlo simulations of
electrically biased short InN and GaN diodes have demonstrated the formation of
stationary one-dimensional gratings of electron concentration
and velocity at nitrogen temperature.\cite{Reggiani-1,Gonzalez-1}

In time (frequency) domain, the cyclic dynamics gives
rise to a resonance phenomenon at the transit-time frequency
$\omega_F = 2 \pi /\tau_F$,  frequently
called optical phonon transient time resonance (OPTTR).
The OPTTR induces a number of interesting effects (for review see
Refs. \onlinecite{Andronov,Reggiani-review}). The most
important manifestation of this phenomenon is the possibility of
microwave amplification and generation in the sub-THz and
THz frequency regions. The microwave generation based on OPTTR
phenomenon was observed experimentally in InP samples
for the frequency range 50 to 300 GHz.\cite{Vorobiev}
For other indicated above  bulk-like materials, the OPTTR
generation was studied theoretically in details.~\cite{Andronov,Reggiani-review}

Discussed cyclic electron dynamics should be even more pronounced
for low-dimensional electron systems. Indeed, for the latter
the onset of the optical phonon emission is much sharper
because of specific features of the density of low-dimensional
states. For comparison, the optical phonon emission rate
$1/\tau_{op}^{em}$ in bulk crystals is proportional to
$\sqrt{E({\bf p})-\hbar \omega_0} \,(0<E({\bf p})-\hbar \omega_0\ll \hbar \omega_0)$,
while for two-dimensional
carriers, $1/\tau_{op}^{em} \propto {\cal H} [E({\bf p})-\hbar \omega_0]$,
where ${\cal H} [x]$ is the Heaviside step-function. In addition
to a sharper threshold of the optical phonon emission,
high mobilities and, thus, quasiballistic motion at energies
$E({\bf p}) < \hbar \omega_0$ can be easily achieved for low-dimensional
electrons, while their concentration can be controlled to avoid
{\em e-e} scattering. These important qualitative conclusions
are supported by papers,~\cite{2DEG-OPTTR,2DEG-OPTTR-2,2DEG-OPTTR-3}
where, by the example of GaN quantum well heterostructures, it
was shown the preference of their use to reach the cyclic
electron motion. Note that  effects similar to OPTTR are predicted
for novel one-dimensional nanostructures - carbon nanotubes.~\cite{streaming-CNT}

Summarizing this short review, one can conclude that
the cyclic electron dynamics is a quite general low-temperature phenomenon characteristic of many polar
materials and heterostructures subjected to an
electric field. It is believed that
the phenomenon may lead to different effects, the most
representative of them are spatial grating of electron
concentration and velocity in finite size diodes and
OPTTR resonance in the time/frequency domain for bulk
samples.

In this paper, by investigating the case of two-dimensional electrons,
we show that  under the conditions of
the cyclic electron motion, a novel type of
weakly damped excitations of drifting electron gas exists.
 These excitations, quite different from well known plasmons, are
periodic in time and in real space oscillations (waves) of
electron concentration and charge both synchronized with
electron redistributions in the momentum space.
Their frequency-wavevector relation has infinite number of
continuous branches, $\omega^k (q)$, with $q$ being the wave vector
of excitations and $k=0, \pm 1, \pm 2...$. The excitation damping
is especially weak or even absent, when the frequency and/or
the wavevector are multiples of $\omega_F$ and/or
$q_F = 2 \pi /l_F$, respectively. Particularly, for $q$ multiple of
$q_F$ standing waves ($\omega$ =0) without any damping appear.

\section{Model and Equations}

It is well established that, for conditions when the cyclic
electron motion can occur, steady state and high frequency
electron transport can be described by semiclassical Boltzmann
transport equation (BTE) for the electron distribution function
in the momentum space. Under conditions (\ref{F-conditions}), the
electron distribution is mainly grouped in the passive region and
essentially elongated along the electric field (the so-called
{\em streaming distribution}). The BTE can be solved
numerically by Monte-Carlo simulations or analyzed analytically
with the use of same approximations adequate to the physical
picture. These approximations include: the Baraff
approximation~\cite{Baraff} to take into account
the anisotropic part and setting to zero the isotropic part of the
distribution function in the active region because of the short
emission time $\tau^{em}_{op}$ (see discussions in Refs.
\onlinecite{Levinson,Gribnikov} for bulk materials, and in
Refs.~\onlinecite{2DEG-OPTTR},  for two-dimensional systems).
In this paper we will use the analytical approach.

Consider a uniform two-dimensional electron gas confined in
narrow quantum well layer at $z=0$ (the quantum well thickness
is negligible in comparison with the characteristic lengths under
consideration). Introduce the reference frame with $X,Y$-axes
in the plane of the quantum well layer and $Z$-axe perpendicular
to this plane. The applied electric field is along the $X$-axe.
Nonequilibrium electrons are described by the distribution
function $\Phi = \Phi ( p_x, p_y, {\bf r}, t)$, which
is, generally, dependent on the momentum, ${\bf p}=\{p_x,p_y \}$,
the coordinate, ${\bf r} =\{x, y\}$, and time, $ t$.
Let $N$ be the areal electron concentration given by doping.
Then we define the distribution function in such a way, that
$N \Phi ({\bf p}, {\bf r}, t) d{\bf p} d{\bf r}$ is the number of electrons
in the phase-space elementary volume $d {\bf p} d {\bf r}$ located at
the phase-space point $\{  {\bf p}, {\bf r} \}$.  The BTE for $\Phi$ is
\begin{equation} \label{BTE-1}
\frac{\partial \Phi}{\partial t}+
\frac{{\bf p}}{m^*}\frac{\partial \Phi}{\partial {\bf r}}-
e {\bf F}  \frac{\partial \Phi}{\partial {\bf p}}=
\hat{L}_{el}\{\Phi\} + \hat{L}_{op}\{\Phi \}\,,
\end{equation}
where the collision integrals
$\hat{L}_{el} \{ \Phi \}$ and $\hat{L}_{op}\{ \Phi \}$ describe
the quasielastic scattering and the inelastic scattering by
optical phonons, respectively.

An excitation of the drifting nonequilibrium electron gas
can be considered as a perturbation of the stationary distribution
with appearance of an additional field. Thus, we present
the distribution function and the field as stationary and
perturbed contributions: $\Phi = \Phi_0 +
\tilde{\Phi}$ and ${\bf F} = {\bf F}_0 + \tilde{\bf F}$, where ${\bf F} = \{- F_0,0,0 \}$
and the field $\tilde{\bf F}$ is defined by the Poisson equation:
\begin{equation} \label{Poisson}
div \tilde{\bf F} = - \frac{4 \pi e N}{\kappa_0} \, \delta [z] \int d {\bf p} \tilde{\Phi}\,,
\end{equation}
where $\kappa_0$ is the dielectric constant of the quantum well surroundings;
$\delta [z]$ is the Dirac function.
The perturbations are assumed to be dependent on time $t$ and
$x$ coordinate as follows:
$  \tilde{\Phi} ({\bf p},x,t) = \tilde{\phi}({\bf p}) \, exp[i q x -i \omega t ] $,
$\tilde{\bf F} (x,z,t) = \tilde{{\bf f}} (z) \, exp[i q x -i \omega t ] $.
In this case the field induced by the perturbed electron gas has
two components: $ \tilde{{\bf f}} (z)=\{{ \tilde{f}}_x (z), 0,  {\tilde{f}}_z (z)\}$.
The Fourier components
$\tilde{\phi}({\bf p}),\, {\tilde{f}}_x (z),  {\tilde{f}}_z (z)$ are dependent on $\omega$ and
$q$.

The BTE in its general form of Eq.~(\ref{BTE-1})
can be simplified by adapting it to the problem under consideration
as follows. Both contributions to the distribution function,
$\Phi_0$ and  $\tilde{\phi}$, are approximated as sums of isotropic
and anisotropic parts:\cite{Baraff}
\begin{eqnarray} {\label{approx-1}}
\begin{array} {c}
\Phi_0 ({\bf p}) = \phi_0 (p) + \phi_1 (p) \,\delta(\theta)\,,\\
\tilde{\phi} ({\bf p}) = \tilde{\phi_0} (p) +\tilde{\phi_1} (p)
\,\delta(\theta),
\end{array}
\end{eqnarray}
where the two-dimensional vector ${\bf p}$ is presented by
its modulus $p$ and the angle $\theta$ to the X-axe.

The next simplifications are the following.
In the passive region, for elastic momentum scattering
it is easy to obtain
$\hat{L}_{el} \{\phi_0\}=\hat{L}_{el}\{\tilde{\phi_0}\} =0$.
While in equations for anisotropic parts, $\phi_1,\tilde{\phi_1}$,
the elastic scattering integrals produce the relaxation terms
$-\{\phi_1,\tilde{\phi_1}\} / \tau_p$. We suppose that the elastic scattering
is due to interaction with acoustic phonons. Then, for two-dimensional
electrons, $\tau_p$ does not depend on the momentum/energy.
In the active region, the collision operator describing
optical phonon emission can be estimated as
$\hat{L}_{op} \Phi \sim \Phi / \tau_{op}^{em}$,
where the optical phonon emission time is the
shortest time in the system ($\tau_{op}^{em} \rightarrow 0)$.
Together with inequality (\ref{necessary-cond}),
this provides the following conditions~\cite{2DEG-OPTTR} for the isotropic parts of
the distribution function: $\phi_0 (p \geq p_0 ), \tilde{\phi_0} (p\geq p_0 )\approx 0$.
In contrast, the anisotropic parts of the distribution functions,
$\phi_1,\tilde{\phi_1} $, are finite at $p=p_0$ due to field induced
electron stream in the momentum space. However, as discussed above, the number
of streaming electrons in the active region is rapidly decreasing because of
the optical phonon emission.
As a result of these simplifications, the BTE is reduced to a system of coupled
ordinary differential equations for $\phi_0, \phi_1, \tilde{\phi_0}, \tilde{\phi_1} $
in the interval $0< p \leq p_0$.

For the steady-state problem these equations are:
\begin{eqnarray} \label{dc-system}
\begin{array} {c}
e  {F_{0}}\frac{d ( p \phi_{1})}{d p}=0 \\
 e F_{0}\left[\pi\frac{d\phi_{0}}{dp}+\frac{1}{p}\frac{d (p \,\phi_{1})}{dp}
\right]=-\frac{\phi_{1}}{\tau_p}
\end{array}
\end{eqnarray}
Two first-order differential Eqs.~(\ref{dc-system}) have to be supplied by
two conditions to determine integration constants. One of these can be obtained from
the normalization of the steady state function. We set $\int \Phi_{0} ({\bf p}) d {\bf p} =1$,  then
we have
\begin{equation} \label{normalization-dc}
\int_{0}^{p_0}  p \,dp \,[2\pi \phi_{0}+\phi_{1}]=1\,.
\end{equation}
The second condition follows from the above discussion:
\begin{equation} \label{bcs-dc-2}
\phi_0 (p_0 )=0\,.
\end{equation}

The spatial and temporal problem is described by equations:
\begin{eqnarray}\label{ac-system}
\begin{array} {l}
i \left[- \omega(2\pi\tilde{\phi_{0}}+\tilde{\phi}_{1})+  \frac{q p}{m^*} \tilde{\phi}_{1} \right]
+ e \frac{F_{0}}{ p}\frac{d ( p \,\tilde{\phi}_{1})}{d p} = 0 \\
i \left[(-\omega - \frac{i}{\tau_p})\tilde{\phi}_{1}+ \frac{q p}{m^*}(\pi\tilde{\phi_{0}}+\tilde{\phi}_{1}) \right]
+ e F_{0} \left[\pi\frac{d \tilde{\phi}_{0}}{d p} \right. \\
\left. +\frac{1}{ p} \frac{d ( p \, \tilde{\phi}_{1}) }{d p}\right]
= e \tilde{f_x} \pi \frac{d {\phi}_{0}}{d p},
\end{array}
\end{eqnarray}
In order to obtain the "boundary" conditions for the latter system, we use the
continuity equation, that follows from the BTE in its general form of
Eq.~(\ref{BTE-1}): $\int d {\bf p} \left[ \frac{\partial \Phi}{\partial t} + \frac{\bf p}{m^*}
\frac{\partial \Phi}{\partial {\bf r}} \right] =0$.
Using our notation, it reads
 \begin{eqnarray} \label{continuity-eq-1}
\int_{0}^{p_0}  p \, dp \,[- \omega (2\pi \tilde{\phi_{0}}+\tilde{\phi_{1}}) +  q \frac{p}{m^*} \tilde{\phi_1}]=0\,.
\end{eqnarray}
Then with the use of the first equation from (\ref{ac-system}), we obtain:
\begin{equation} \label{bcs-a-c-1}
p \tilde{\phi}_1 \large|_{p \rightarrow 0} - p_0 \tilde{\phi}_1 (p_0 ) =0\,.
\end{equation}
The second boundary condition is similar to Eq.~(\ref{bcs-dc-2}):
\begin{equation} \label{bcs-a-c-2}
 \tilde{\phi_0} (p_0 ) = 0\,.
\end{equation}

The {\em ac}-electric field component, $\tilde{f}_x$, that enters into the right hand side
of the second equation of the system (\ref{ac-system}), is defined
via the Fourier component of the electrostatic potential, $\tilde \varphi$:
  $\tilde{f}_x = - i q \tilde \varphi$. The latter obeys the Poisson equation:
\begin{equation} \label{Poisson-2}
\frac{d^2 \tilde{\varphi}}{d z^2} - q^2 \tilde{ \varphi} = \frac{4 \pi e N}{\kappa_0}\, \delta (z) \int_0^{p_0} dp\,p
[2 \pi \tilde{\phi}_0 +\tilde{\phi}_1]\,
\end{equation}
with the boundary conditions $\tilde{ \varphi} (z \rightarrow \pm \infty)=0$.
These relations finalize the mathematical formulation of the steady-state and
wave-like excitation problems for the streaming two-dimensional electron gas.

\section{Solutions of equations}
To proceed further it is convenient to introduce the following dimensionless
variables:
\begin{equation} \label{dimensionless}
X = \frac{x}{l_F}\,,T = \frac{t}{\tau_F}\,,\,P = \frac{p}{p_0}\,, \,\Omega = \omega \tau_F\,,\, \Gamma= \frac{\tau_F}{\tau_p}\,,\,Q = q l_F.
\end{equation}
We substitute $p_0^2 \phi_0 \rightarrow  \phi_0\,,\,p_0^2\tilde{ \phi_0} \rightarrow \tilde{ \phi_0}$, etc.
Thus, the functions $\phi_0,\,\phi_1,\,\tilde{\phi_0},\,\tilde{\phi_1}$ are now dimensionless.

\subsection{ Solutions for steady state problem}

The solutions of Eqs.~(\ref{dc-system})-(\ref{bcs-dc-2}) for the steady state
problem  are
\begin{eqnarray} \label{dc-solutions}
\phi_0 (P) = \frac{- 2 \Gamma}{\pi \left[ 2+ \Gamma  \right]}  \ln P \,,\\
\phi_1 (P) = \frac{2}{  \left[  2  + \Gamma \right]} \frac{1}{P}\,
\end{eqnarray}
 Eqs. (\ref{dc-solutions}) describe
 the electron distribution  quite well almost in the whole momentum
 space, except a narrow interval corresponding to the electron penetration into the active
 region given  by Eq.~(\ref{penetration}) and the respective interval nearby $p=0$.~\cite{2DEG-OPTTR}

Having $f_{0}$ and $f_{1}$ one can calculate the average electron energy
and the drift velocity:
\begin{eqnarray}
\epsilon_{av} = \int d {\bf p} \frac{{p}^2}{2 m^*} \Phi_{0} ({\bf p}) =
\frac{8 + 3 \Gamma }{12\,[2 + \Gamma]}  \,\hbar \omega \,,\\
v_{dr} =  \int d {\bf p} \frac{p_x}{ m^*} \Phi _{0}({\bf p}) =
\frac{1}{[2 + \Gamma ]}\,\frac{p_0}{m^*}\,, \label{v-drift}
\end{eqnarray}
For the case of the strong inequality (\ref{F-conditions}), we obtain the characteristics of
the "ideal" electron streaming regime:  $\epsilon_{av} = \hbar \omega_0/3,\,v_{dr} = p_0/2 m^* $.

\subsection{Solutions for the space and time dependent problem}

Using the dimensionless variables of Eqs.~(\ref{dimensionless}) one can rewrite
equations Eqs.~(\ref{ac-system}) in the form:
\begin{eqnarray} \label{ac-system-dimensionless}
\begin{array} {l}
i \left[ - \Omega \left( 2 \pi P \tilde{\phi}_0 +P \tilde{\phi}_1 \right) + 2 QP^{2} \tilde{\phi}_1 \right] +
\frac{d \left(P \tilde{\phi}_1  \right)}{d P} =0\,,\\
i \left[  \left(- \Omega - i \Gamma \right) P \tilde{\phi}_1 + 2 Q P \left( \pi P \tilde{\phi}_0
+ P \tilde{\phi}_1 \right) \right]\\
 + \left[ \pi \frac{d \left(P \tilde{\phi}_0 \right)}{d P}
 -\pi\tilde{\phi}_0 +
 \frac{d \left( P \tilde{\phi}_1 \right)}{d P} \right] = \pi \frac{\tilde{f}_x}{F_0} P \frac{d \phi_0}{d P}\,.
\end{array}
\end{eqnarray}
The following substitutions
\begin{eqnarray} \label{substitution}
\begin{array} {c}
\pi P \tilde{\phi}_0 = \chi_0 \exp \left[ i \left( \Omega P - Q P^2 \right) \right]\,,\\
P \tilde{\phi}_1 = \chi_1 \exp \left[ i \left( \Omega P - Q P^2 \right) \right] \,,
\end{array}
\end{eqnarray}
transform Eqs~(\ref{ac-system-dimensionless}) to the simpler form
\begin{eqnarray} \label{ac-system-simplified}
\begin{array} {c}
\frac{d \chi_1}{d P} = 2 i \Omega \chi_0\,,\\
\frac{d \chi_0}{d P} + \left[3 i \Omega - \frac{1}{P} \right] \chi_0 + \Gamma \chi_1
=   \pi \frac{\tilde{f}_x}{F_0} P \frac{d \phi_0}{d P} \\
 \times \exp \left[- i \left( \Omega P - Q P^2 \right) \right] \,.
\end{array}
\end{eqnarray}
The latter system of two equations can be reduced to a single equation for the
function, $\chi_1 (P)$, that we present as
\begin{equation} \label{basic-equation}
\frac{d^2 \chi_1}{d P^2} + \left[b -\frac{1}{P} \right] \frac{d \chi_1}{d P} + c \chi_1
= \frac{{\tilde {f}}_x}{F_0} R (P\, | \Omega, Q)\,,
\end{equation}
where we designate
\begin{eqnarray} \label{parameters}
\begin{array} {c}
b = 3 i \Omega,\,  c = 2 i \Omega \Gamma,\, \\
R (P\,| \Omega, Q) =  2 \pi i \Omega P \frac{d \phi_0}{d P}
\exp \left[- i \left( \Omega P - Q P^2 \right) \right]\,.
\end{array}
\end{eqnarray}
Note that, in the left hand side of Eq.~(\ref{basic-equation}), the parameters, $b,\,c$,
are dependent only on $\Omega$, while the right hand side parametrically depends on
both $\Omega$ and $Q$. From Eqs.~(\ref{bcs-a-c-1}), (\ref{bcs-a-c-2}) and the first
equation of the system (\ref{ac-system-simplified}) it follows the boundary conditions
to Eq.~(\ref{basic-equation}):
\begin{eqnarray} \label{bcs-basic}
\begin{array} {c}
\chi_1 (0) = \chi_1 (1) \exp \left[ i \left( \Omega  - Q \right) \right]\,,\\
\frac{d \chi_{1} }{d P} (1)=0\,.
\end{array}
\end{eqnarray}

The general solution to Eq.~(\ref{basic-equation}) can be presented as
\begin{eqnarray}\label{ac_solution}
\chi_{1} (P) =C_{1} \Psi_{1} (P) + C_{2} \Psi_{2} (P) + \nonumber \\
\frac{\tilde{f}_{x}}{F_0} \left[  \Psi_{1} (P)
 \int_{P}^{1} \frac{R(P') \Psi_{2} (P')}{W_{\Omega}(P')} dP'  -
 \Psi_{2} (P) \int_{P}^{1} \frac{R(P') \Psi_{1} (P')}{W_{\Omega}(P')}d P' \right]\,
\end{eqnarray}
where $\Psi_1 (P),\,\Psi_2 (P)$ are two independent solutions of the congruent
homogeneous equation, $W_{\Omega} (P)$
is the Wronskian of this equation, $C_1,\,C_2$ are arbitrary constants.
The functions $\Psi_{1} (P)$ and $\Psi_{2} (P)$ can be expressed via the Kummer functions
of first and  second kinds,~\cite{Kummer-function} $M (u,w,z)$ and $U(u,w,z)$, respectively:
\begin{eqnarray}\label{fundfunction}
\Psi_{1} (P )=P^{2} \exp(-\alpha P) M [\beta,3,\epsilon P]\,,\, \nonumber\\
\Psi_{2} (P )=P^{2} \exp(-\alpha P) U [\beta,3,\epsilon P]\,,\nonumber
\end{eqnarray}
with $\epsilon = \sqrt{b^2- 4 c}\,,\,\alpha = (b + \epsilon )/2$ and
$\beta = 3/2 - b/2 \epsilon$.

 For the problem under consideration, the constants $C_1$ and $C_2$ should be found
 by using the boundary conditions (\ref{bcs-basic}). The latter result in the following
equations:
\begin{eqnarray}
C_{1}\frac{d\Psi_{1}}{d P}(1)+C_{2}\frac{d\Psi_{2}}{dP}(1)&=&0   \label{system-C1-C2-1} \\
C_{1}\left\{\Psi_{1}(0)-\Psi_{1}(1) \exp\left[ i \left( \Omega - Q \right) \right] \right\}+
 C_{2}\left\{\Psi_{2}(0)-\Psi_{2}(1)\exp {\left[ i \left( \Omega - Q \right) \right]} \right\}&=&\nonumber \\
-\frac{{\tilde{f}}_x}{F_0} \left[ \Psi_{1}(0)J_{2}(0)-\Psi_{2}(0)J_{1}(0) \right]\,, \label{system-C1-C2-2}
\end{eqnarray}
where  $J_{1,2} (P) \equiv \int_{P}^{1} \frac{R (P') \Psi_{1,2} (P')}{W_{\Omega} (P')} d P'$.
The {\em ac} electric field, $\tilde{f}_x$, that appears in the right hand side of
Eq.~(\ref{system-C1-C2-2}), is defined by the Poisson Eq.~(\ref{Poisson-2}):
\begin{eqnarray} \label{f_x}
&&\frac{\tilde{f}_x}{F_0} =  sgn(Q)  \frac{2 \pi i e N}{\kappa_0 F_0}\, \int_0^1 dP P \left[ 2 \pi \tilde{\phi}_0 +
  \tilde{\phi}_1 \right] = \nonumber \\
&&  sgn(Q)  \frac{2 \pi i e N}{\kappa_0 F_0} \,\int_0^1 d P \left[ \chi_1- \frac{i}{\Omega} \frac{d \chi_1}{d P}\right]
 \,exp \left [ i \left( \Omega P - Q P^2 \right) \right]\,,
\end{eqnarray}
where $sgn(Q)$ stands for sign of Q. Substituting (\ref{ac_solution}) to Eq.~(\ref{f_x}) one can find $\tilde{f}_x$
as a linear homogeneous form of the constants $C_1, C_2$:
\begin{equation} \label{linear-form}
\frac{\tilde{f}_x}{F_0} = {\cal N} \left(  A_1  C_1 + A_2 C_2 \right)\,,
\end{equation}
where $ A_{1,2} = A_{1,2} \{\Psi_1,\Psi_2 |\Omega,Q \}$  are expressed via bulky integrals  from
some combinations of the functions $\Psi_1 $ and $\Psi_2$.
We introduce the parameter
\begin{equation} \label{L}
{\cal N} = \frac{N}{N_{ch}}\,, \,\,N_{ch} \equiv \frac{\kappa_0 F_0}{2 \pi e} \,,
\end{equation}
which has the meaning of a dimensionless electron concentration.
From the structure of Eq.~(\ref{system-C1-C2-2}) it follows that this parameter
determines the coupling of the dynamics of individual electrons and their collective motion arising due to
the self-consistent electric field, ${\bf \tilde{f}}$.
Eqs.~(\ref{system-C1-C2-1}), (\ref{system-C1-C2-2}) supplemented by the relationship (\ref{linear-form})
compose the system of homogeneous linear algebraic equations for the constants $C_1,\,C_2$.
The solvability condition of this system is determined by the zero of the associated determinant,  $\Delta (\Omega, Q) =0$.
The latter condition gives the dispersion relation, $\Omega (Q)$, for required excitations of
the drifted electron gas.

\subsection{Frequency and wavevector dispersion of electron conductivity and dielectric permittivity
of the electrons.}

The solvability of the obtained equations can be also presented in a more
usual form. Indeed, the high frequency current can be calculated as
\begin{equation} \label{h-f-current}
\tilde{j}_x = -\frac{e N p_0}{m^*} \int_0^1 dP P^2 \tilde{\phi}_1 (P) \,,
\end{equation}
where parametrically dependent on $\Omega$ and $Q$ function $\tilde{\phi}_1 (P)$ is the
solution of Eqs.~(\ref{substitution}), (\ref{ac_solution}), (\ref{system-C1-C2-1}),
(\ref{f_x}) obtained at a given $\tilde{f_x}/F_0$.
Then, the high frequency conductivity is
\begin{eqnarray} \label{h-f-conductivity}
\sigma \left( \Omega, Q \right) = \frac{\tilde{j}_x}{\tilde{f}_x} =\, \frac{e N p_0}{m F_0} \,
\nu\left(\Omega, Q \right) \equiv e \mu_0 N \frac{F_{cr}}{F_0} \nu\left(\Omega, Q \right) \,,\nonumber \\
 \nu \left(\Omega, Q \right) \equiv -\frac{F_0}{\tilde{f}_x} \int_0^1 dP P^2 \tilde{\phi}_1 (P)\,, \label{h-f-conductivity}
 \end{eqnarray}
 where $\mu_0 = e \tau_p/m$ is the low field mobility, $\nu \left(\Omega, Q \right)$ does not depend on
 $\tilde{f}_x/F_0$ and has a meaning of dimensionless electron mobility dependent on both the frequency
 and  the wavevector.

With the help of Eq.~(\ref{h-f-conductivity}), one can calculate the
dielectric permittivity of the electrons:
\begin{equation} \label{epsilon}
\epsilon \left( \Omega, Q \right) = 1 + i 2  {\cal N} \frac{|Q|}{\Omega}\, \nu \left(\Omega, Q \right) \,.
\end{equation}
It can be easily proved that the usual requirement
\begin{equation} \label{epsilon=0}
\epsilon{\left( \Omega, Q \right)} =0\,,
\end{equation}
that defines the possible {\em collective} excitation modes of the electron gas, is identical to the
above obtained solvability condition.

\begin{figure}
\includegraphics[height=9cm,width=9cm]{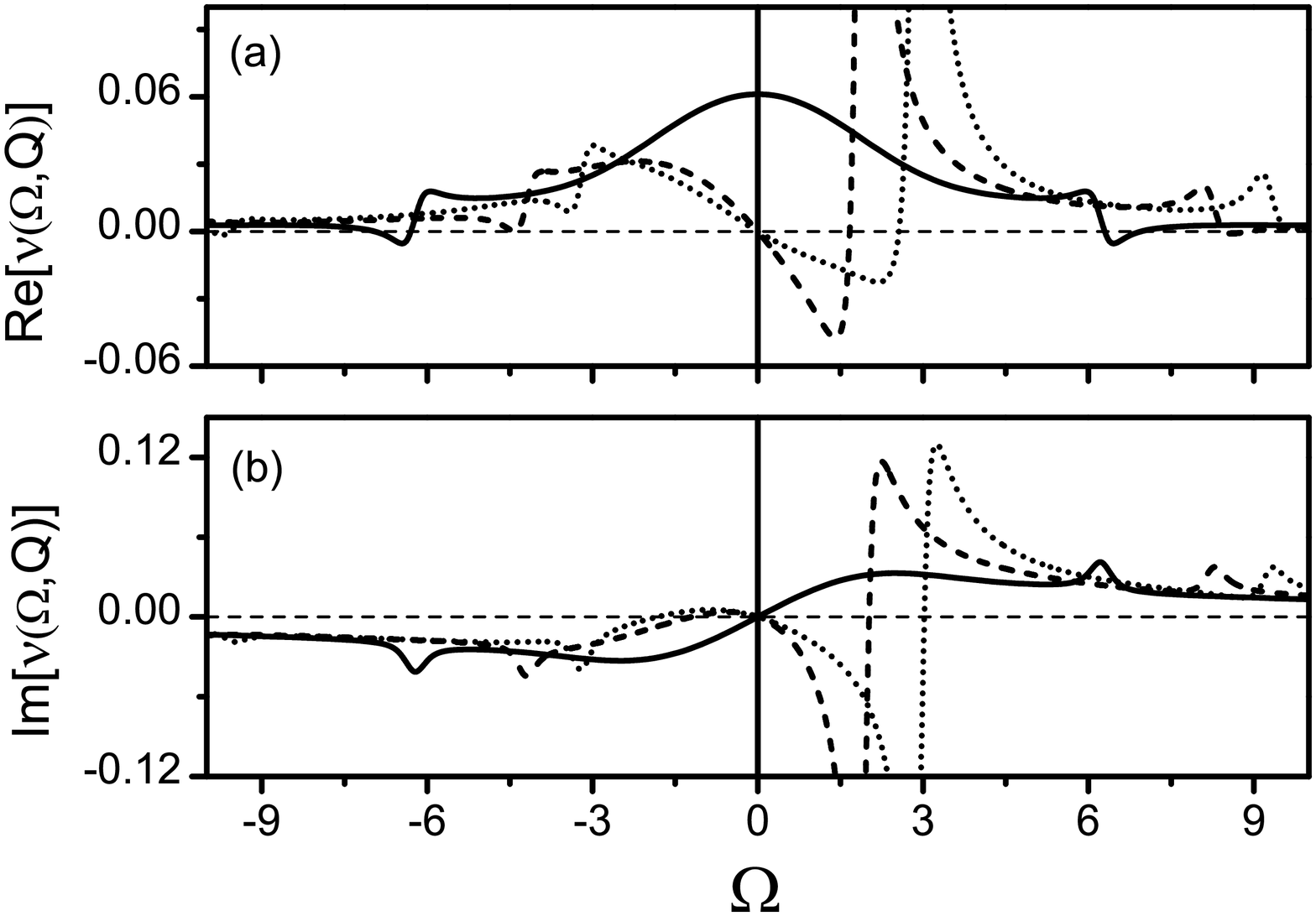}
\caption{ Real (a) and imaginary (b) parts of the dimensionless high frequency mobility of
Eq.~(\ref{h-f-conductivity}) as functions of $\Omega$ for different $Q$.
Results are presented for $F_0/F_{cr} =3 $. Full, dashed and dotted
lines correspond to $Q=0,\,1,\,2$, respectively. }
\label{fig-sigma}
\end{figure}

For what follows, it is useful to discuss briefly some properties of the high frequency mobility,
$\nu (\Omega, q)$. Fig.~\ref{fig-sigma} shows the results of  calculations of
$\nu (\Omega, Q)$ for $F_0/F_{cr} =3$.  The OPTTR  becomes apparent near
$\Omega = \pm 2\pi$. Indeed, one can see typical frequency dispersion
of the conductivity in narrow frequency intervals and, particularly, a change of the sign of
the dynamic conductivity ($Re[\nu] < 0$). The latter feature has been in focus of
previous studies.~\cite{Reggiani-1,Andronov,Reggiani-review,2DEG-OPTTR}
At $Q \neq 0$,  drift of the electrons is revealed: the characteristic
feature arises at $\Omega \approx V_{dr} Q$,
where $V_{dr}$ is the dimensionless drift velocity defined as $V_{dr} = 2 v_{dr} m/ p_0$ with the field dependent $v_{dr}$ given by Eq.~(\ref{v-drift}). Besides, a negative dynamic conductivity
occurs at the Cerenkov region, $0< \Omega < V_{dr} Q$, while the OPTTR frequencies
are shifted by factor $V_{dr} Q$.
Note, the absolute values of $\nu (\Omega, Q)$ are relatively small. This is typical for
conditions of the streaming  regime, when the drift velocity saturates and
the differential mobility tends to zero.

\section{Solutions at ${\cal N} \rightarrow 0$.}

First, we study the space and time dependent solutions
neglecting self-consistent electric field  ${\bf \tilde{f}}$. According to the relationship
(\ref{linear-form}), this corresponds to the limiting case
\begin{equation} \label{L}
{\cal N} \ll 1 \,.
\end{equation}
In this limit, the electron gas excitations can be thought as {\em phased}
motion of the electrons in the real space and the momentum space
under the conditions of cyclic dynamics of individual electrons. Importantly,
all results for this limit depend on the single parameter, $F_0/F_{cr} = \tau_p/\tau_F = 1/\Gamma$.

For such a case, from Eqs.~(\ref{system-C1-C2-1}) and (\ref{system-C1-C2-2})
we immediately obtain the condition of solvability of these equations in the form
\begin{equation} \label{Det0}
\Delta_0 \left( \Omega, Q \right) = \Psi_{1} (0) \frac{d \Psi_{2}}{d \rho} (1) - \Psi_{2} (0)
\frac{d \Psi_{1}}{d \rho}(1) - W_{\Omega}(1) \exp \left[ i \left( \Omega - Q \right) \right] = 0\,,
\end{equation}
that can be rewritten as
\begin{equation} \label{Det0-1}
\exp \left (-i Q \right) = \frac{\Psi_{1}(0)\frac{d\Psi_{2}}{d\rho}(1)-\Psi_{2}(0)
\frac{d\Psi_{1}}{d\rho}(1)}{W_{\Omega}(1)} \exp \left( -i \Omega \right)
\equiv {\cal S} (\Omega^{\prime},\Omega^{\prime\prime}|F_{0})\,.
\end{equation}
In the latter form, the left hand side depends  on $Q$, while the right hand side is a function of
$\Omega$. We will look for wave-like excitations with a real wavevector, $Q$, while
the frequency can be a complex value, $\Omega = \Omega' + i \Omega''$.  To find $\Omega'$
and $\Omega''$ at a given $Q$ we have two equations
\begin{eqnarray}
abs \left[ {\cal S} (\Omega^{\prime},\Omega'') \right]=1 \,,\label{Kampen1} \\
Q= - arg \left[ {\cal S} (\Omega^{\prime},\Omega'' )  \right]\, \pm 2 \pi k  \,, \label{Kampen2}
\end{eqnarray}
where $k$ is an integer, $k=0,1,2...$. Here we introduce the absolute value, $abs \left[ {\cal S} \right]$,
and the phase, $arg \left[ {\cal S} \right]$, of the complex function $ {\cal S} (\Omega^{\prime},\Omega'' ) $.
Eq.~(\ref{Kampen1}) defines the damping  of the excitations, $\Omega^{{\prime}{\prime}}$,
as a function of the frequency $\Omega'$. Then, Eq.~(\ref{Kampen2}) defines
implicitly  the dispersion relation of sought-for wave excitations,
$\Omega^{\prime}(Q)$ for the limiting case of Eq.~(\ref{L}).

\begin{figure}
\includegraphics[height=9cm,width=12cm]{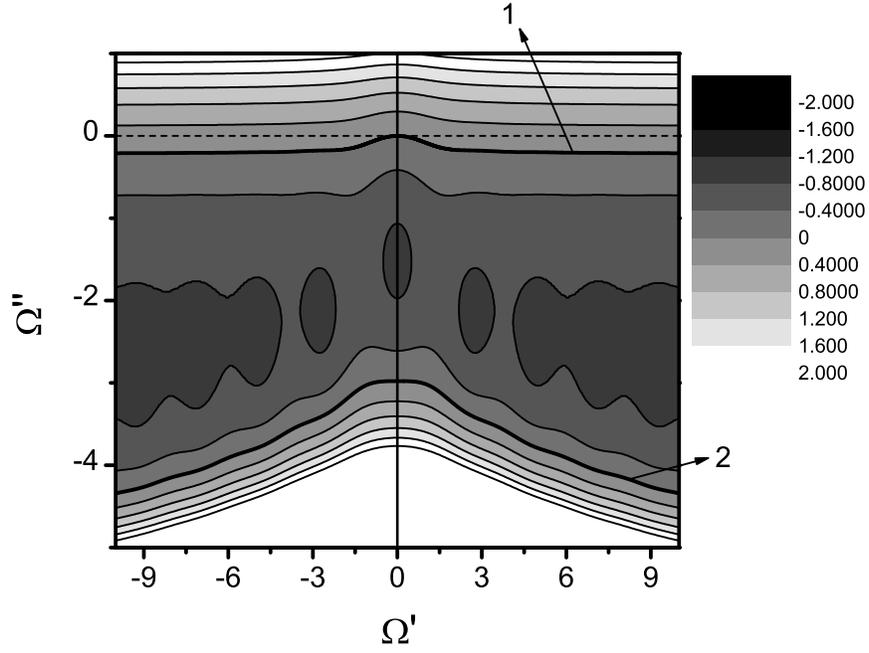}
\caption{Density plot of the function ${\cal G} (\Omega', \Omega")$. Curves 1 and 2 correspond to
${\cal G}=0$ and present  $\Omega''(\Omega')$-dependencies, i.e., the damping
of the excitations. }
\label{fig-absG}
\end{figure}

To understand the nature of the solutions of Eq.~(\ref{Kampen1}), one should analyze the
following function of two variables ${\cal G} (\Omega', \Omega^{\prime \prime}) =
abs \left[ {\cal S} (\Omega', \Omega^{\prime \prime}) \right] -1$.
This function parametrically depends on a single value: the dimensionless field $F_0/F_{cr}$.
In Fig.~\ref{fig-absG}, a  typical density plot of ${\cal G} (\Omega', \Omega^{\prime \prime})$  is presented for
$F_0/F_{cr}=3$ (see below discussion of actual parameters for particular materials).
As seen from this figure, there are two kinds of solutions, $\Omega''= \Omega'' \left[\Omega' \right]$.
The first is characterized by small $\Omega''$ and, particularly, $\Omega'' (\Omega'=0) =0$. The second
kind of solutions gives $\Omega''$ of the order of $\Omega'$. Below we will concentrate on the first
kind of solutions, because the electron gas excitations corresponding to such solutions are weakly damped.

\begin{figure}
\includegraphics[height=9cm,width=9cm]{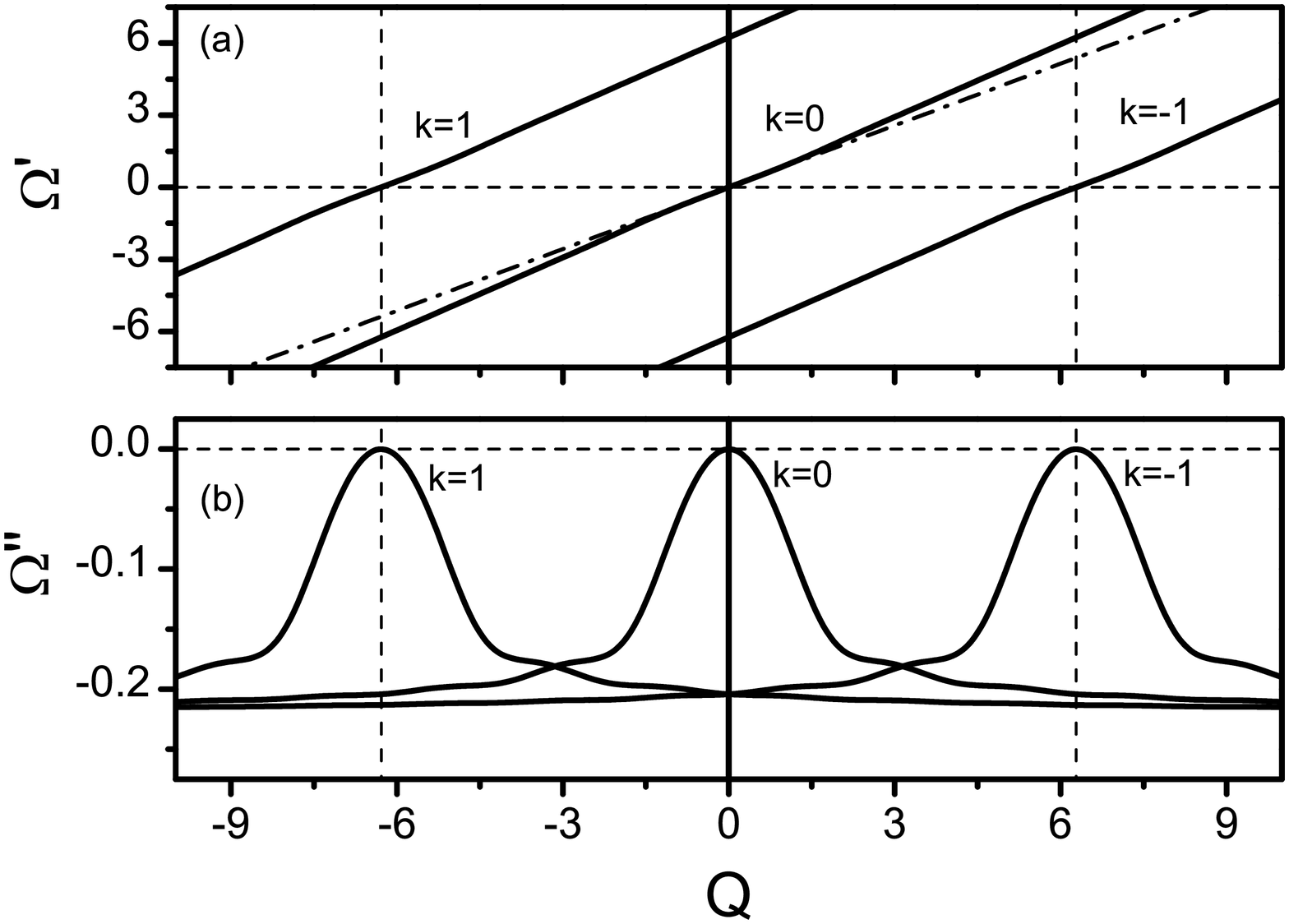}
\caption{Real (a) and imaginary  (b) parts of the frequency as functions
of the wavevector found from Eqs.~(\ref{Kampen1}), (\ref{Kampen2})
for three branches of the dispersion relation at $F_{0}/ F_{cr}= 3 $. Dash-dotted line corresponds to $\Omega^{\prime}=V_{dr}Q$.
The vertical dashed lines mark wavevectors equal to  $Q= \pm 2 \pi$.
Note that the vertical scales in (a) and (b) are very different.}
\label{set-branches-1}
\end{figure}

From Eq.~(\ref{Kampen2}), it follows that there is an infinite set of
branches of the dispersion relation  corresponding to different values of the integer $k$.
We will call the branch of $k=0$ as the acoustic-like dispersion branch. From Eqs.~(\ref{Kampen1}) and
(\ref{Kampen2}), for the acoustic branch one can find a function $\Omega = \Omega^0 (Q)$. Then, any
other 'satellite' branch can be obtained by a simple translation in the $\{\Omega, Q\}$-plane  along
the $Q$-axe:  $\Omega = \Omega^k (Q) =   \Omega^0 (Q \pm 2 \pi k)$.  The "basic vector"
of translation  is $Q_{0} = 2 \pi$. Its dimension value is $2 \pi/l_F$. In the real space, this value
corresponds to a wavelength of the excitation strictly equal to the characteristic
spatial period of the cyclic electron motion, $l_F$.
For the acoustic branch, we find that $\Omega^0 \approx V_{dr }Q$ at $Q \rightarrow 0$.
At finite $Q$, the acoustic branch can be obtained numerically: the result is shown
in Fig.~\ref{set-branches-1} (a). In this figure, we presented also two satellite branches with $k=-1,1$.
The latter branches have finite frequencies at $Q \rightarrow 0$, i.e., they
can be thought as of optical type.
Interestingly, the frequencies of the satellite branches at finite wavevectors $Q= Q_0 k$
are {\em exactly} zero. This means that the corresponding electron gas excitations
are {\em standing} waves. For convenience in Fig.~\ref{set-branches-1} (b), we
show also the imaginary part of the frequency, $\Omega^{\prime \prime} (Q)$, for each of the branches.
The absolute values of $\Omega^{\prime \prime}$ are small in comparison with $\Omega'$.~\cite{remark-1}
The solutions corresponding to the standing waves have zero
damping, $\Omega^{\prime \prime} (k  Q_0) =0$.

For the latter particular case ($\Omega = 0$,  while $\Gamma \neq 0$),
the solutions can be found in an explicit form.   Indeed, from
Eqs.~(\ref{ac-system-simplified}) we find $\chi_1 = C =constant\,,\,\chi_0 = - C \Gamma P ln P$.
Then, from the boundary conditions  (\ref{bcs-basic})
we find the allowed $Q = 2 \pi k\,,k \neq 0$. According to Eqs~(\ref{substitution}),
contributions to the distribution function are
\begin{eqnarray} \label{st-waves}
\tilde{\phi_0} = - \frac{C \Gamma}{\pi } ln P \times  exp(-i 2 \pi k P^2)\,,\\
\tilde{\phi_1} = \frac{C}{P} \times exp(-i 2 \pi k P^2)\,,
\end{eqnarray}
Interestingly, the Fourier component of the perturbed electron density corresponding to
these solutions is not zero, while the perturbation of the average electron velocity is zero, i.e. there are
no electron fluxes in real space.

Additional analytical results can be obtained for the ultimately simple case
adequate to $\Gamma =0$ (absence of any scattering in the passive region). In this limit,
from Eqs.~(\ref{ac-system-simplified}),  we obtain $\chi_0 =0,\,\chi_1 = C =constant$. The corresponding
contributions to the distribution function are:
$$
\tilde{\phi}_0 = 0,\,  \tilde{\phi}_1 = \frac{C}{P} exp[i(\Omega P - Q P^2)]\,.
$$
Then, from the boundary conditions (\ref{bcs-basic}) we find $\exp[ i(\Omega - Q)] =1$,
i.e.,  the dispersion relation gives an infinite set of straight lines:
$\Omega^k_{\Gamma=0} (Q) = Q + 2 \pi k\,, k=0,\, \pm 1, ... \,\,\,(\Gamma =0)\,.$
At nonzero but small $\Gamma$, this dispersion relation can be corrected by using
the perturbation method: $\Omega = \Omega^k_{\Gamma=0} + \Gamma
{\cal B} (Q+ 2 \pi k)$, where the function ${\cal B} (X)$ is defined as
$$
{\cal B} (X) =2 X \int_0^1 dx x e^{-i 3Xx} \int_1^x \frac{dy}{y} e^{i 3 X y} \,.
 $$
Near the points $Q= - 2 \pi k$, where $\Omega^k_{\Gamma=0}$ is small, we obtain approximately
$$
\Omega \approx (Q+ 2 \pi k)  \left(1 - \frac{\Gamma}{2} \right)  -
i \frac{\Gamma}{3}  {(Q+ 2 \pi k)^2}\,.
$$
Thus, electron momentum dissipation in the passive region leads to the damping of
the wave excitations, as well as to renormalization of the excitation frequency. However, the standing waves
are not damped.

The found dispersion relation and solutions can be explained by time-dependent
{\em phased}   motion of many electrons in both the real space and the momentum space.
The properties of such a phased motion are very distinct for different $Q$ and $k$. Indeed,
for the acoustic branch at $\Omega,\,Q \rightarrow 0$ the solutions correspond to the simple
drift of a perturbed electron concentration with velocity $V_{dr}$ (this type of solutions
can be obtained, particularly, in the simplest hydrodynamical
models).  At $k \neq 0$ and small $Q$  (optical-like branches), the time-dependent
phased motion occurs mainly in the momentum space. At $k \neq 0,\,Q \approx k Q_0
$, the electron motion in the real space  is phased with that in the momentum space, but it is
time independent.

\section{Solutions at ${\cal N}$ of a finite value.}

For the case of a finite value of ${\cal N}$,
when collective charge effects are involved, we performed a numerical analysis of the
equations. Before presenting this analysis, we note that, due to the scaling given by
Eqs.~(\ref{dimensionless}), the results obtained for
${\cal N}=0$ were dependent  only on the dimensionless parameter $F_0/F_{cr}$.
Now, for a given ${\cal N}$   we calculate $\nu (\Omega, Q)$ and $\epsilon(\Omega,Q)$
according to Eqs.~(\ref{h-f-conductivity}), (\ref{epsilon}),    then we solve
Eq.~(\ref{epsilon=0}) for the complex frequency $\Omega$ at a given real wavevector $Q$.
Representative examples of these calculations are given in Fig.~\ref{Omega-Q-with charge-1},
where the real and imaginary parts of $\Omega$ are shown as functions of $Q$ for three
branches of excitation spectra at $F_0/F_{cr} =3$ and ${\cal N} =7.4$.
Comparison of the results for the real part of the frequency, $\Omega'$,
with those presented in Fig.~\ref{set-branches-1} shows that, at a finite ${\cal N}$, noticeable
changes occur for the acoustic branch of the excitation spectra, while changes of the
optical-like branches ($k=\pm 1$) are much smaller. Particularly, there is no effect of
${\cal N}$ on the existence of the standing waves ($\Omega = 0$), which have the same
spatial periods ($Q = 2 \pi k$). A more detailed evolution of the acoustic-like branch when
${\cal N}$  varies is
illustrated in Fig.~\ref{main-branch} (a). Comparing the damping of the
excitations, $\Omega''$, presented in Fig.~\ref{Omega-Q-with charge-1} (b) one can make similar conclusion: at finite values of ${\cal N}$ the damping
of the acoustic-like excitation branch increases  significantly, while that of
the optical-like branches is less affected. Specifically, the standing waves remain strictly
undamped.  In  Fig.~\ref{main-branch} (b) the damping of the acoustic-like branch is
presented for different ${\cal N}$.

Remarkably, the solutions for the standing waves can be obtained {\em exactly} for
arbitrary ${\cal N}$. The solutions include, particularly, the expression for the amplitude of
the  electric field, ${\tilde f}_x$, associated with space and time dependent redistribution
of the charged carriers. These solutions are:
\begin{eqnarray}
\Omega=0\,, Q= 2 \pi k \,,\nonumber \\
\tilde{\phi_0} = - \frac{C \Gamma}{\pi } \,
\frac{2+\Gamma+2 i \Gamma {\cal N} I_1(k)}{2+\Gamma+4i\Gamma {\cal N} I_2 ( k)}
\,ln P \times  exp(-i 2 \pi k P^2)\,, \nonumber  \\
\tilde{\phi_1} = \frac{C}{P} \times exp(-i 2 \pi k P^2) \,,  \label{st-waves-2} \\
\frac{\tilde{f_x}}{F_0} = i {\cal N} C \, \frac{[2+\Gamma]
\left[ I_1 (k) -2 \Gamma I_2 (k) \right]}{2 + \Gamma +4 i \Gamma {\cal N} I_2( k)}\,,\nonumber \\
\end{eqnarray}
Here $C$ is an arbitrary constant, functions $I_1(k)$ and $I_2(k)$ are
$$ I_1 (k) = \int_0^1 dP  e^{ -i 2 \pi k P^2} \,,\,\, I_2 (k) = \int_0^1 dP P\, lnP \,e^{ -i 2 \pi k P^2} \,. $$

As indicated above, the found solutions correspond to a time-dependent motion
of the electrons in real space highly correlated with that in momentum space.
Spatial charge effects are related to electron behavior in real space.
To analyze them, one can introduce the concentration
of the "isotropic" electrons, $\tilde N_0$, and the concentration of
the "streaming" electrons, $\tilde N_1$, which are determined through the disturbed
distributions $\tilde \phi_0$ and $\tilde \phi_1$, respectively. The total perturbed
concentration of the excitation wave is $\tilde N = \tilde N_0 + \tilde N_1$.
To understand the main characteristics of the time and real space dependences of these concentrations
we neglect the small damping of the excitations, so that the dependences are
\begin{eqnarray} \label{N0-N1}
\begin{array} {c}
\tilde N_0 (X,T) = \tilde N_0 (Q, k) \,\cos[ \Omega^{\prime k} (Q) T - Q X + \delta_0]\,,\,\\
\tilde N_1 (X,T) = \tilde N_1  (Q, k) \,\cos[ \Omega^{\prime k} (Q) T - Q X + \delta_1]\,.
\end{array}
\end{eqnarray}
Here we take into account that a given spectrum branch number, $k$, and
wavevector, $Q$, determine the excitation frequency, $\Omega^k (Q)$,  and the particular
solutions. Then, $\delta_0$ and $\delta_1$ are phase shifts dependent on $Q$ and  $k$.

\begin{figure}
\includegraphics[height=8cm,width=8cm]{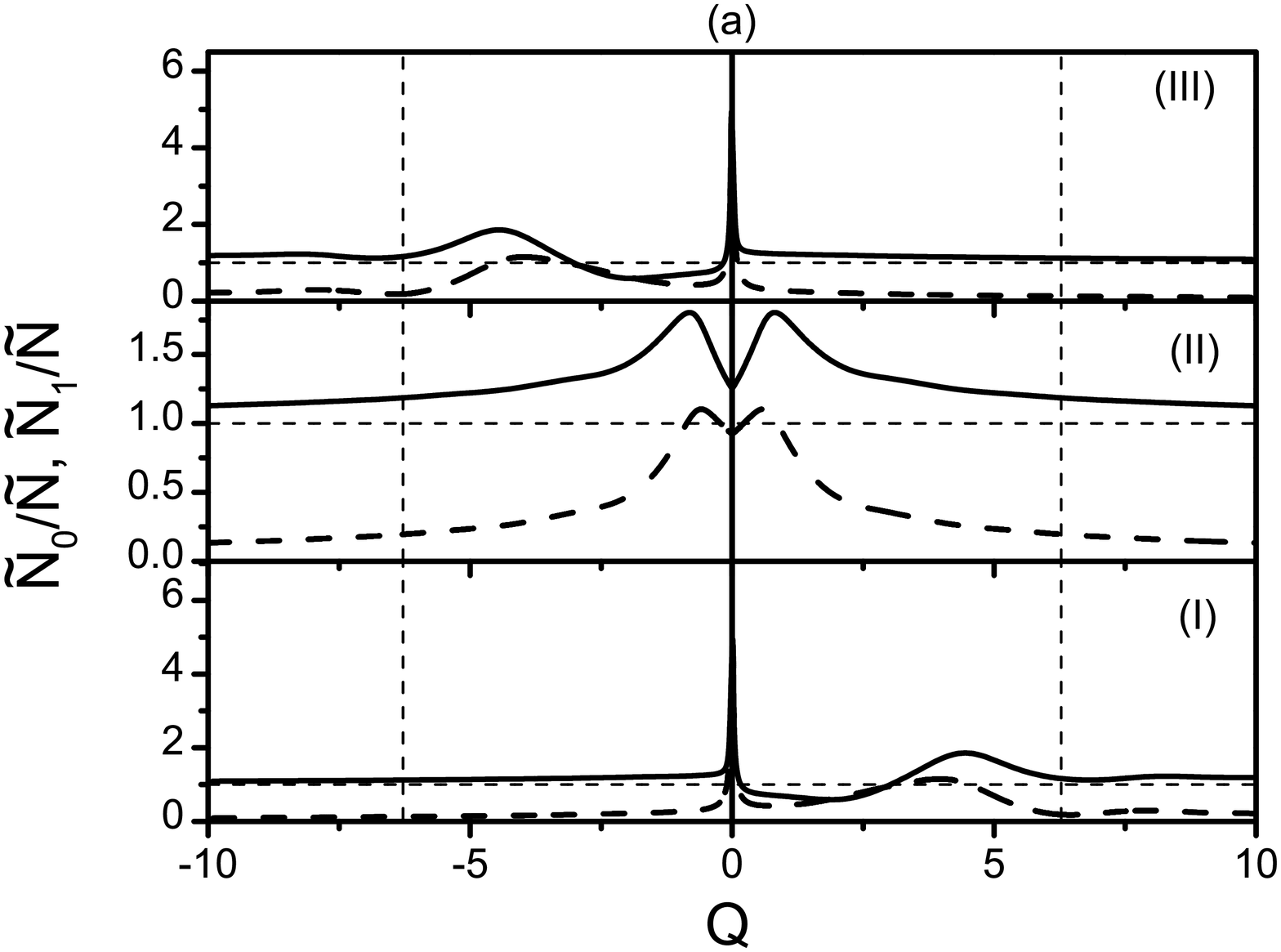}
\includegraphics[height=8cm,width=8cm]{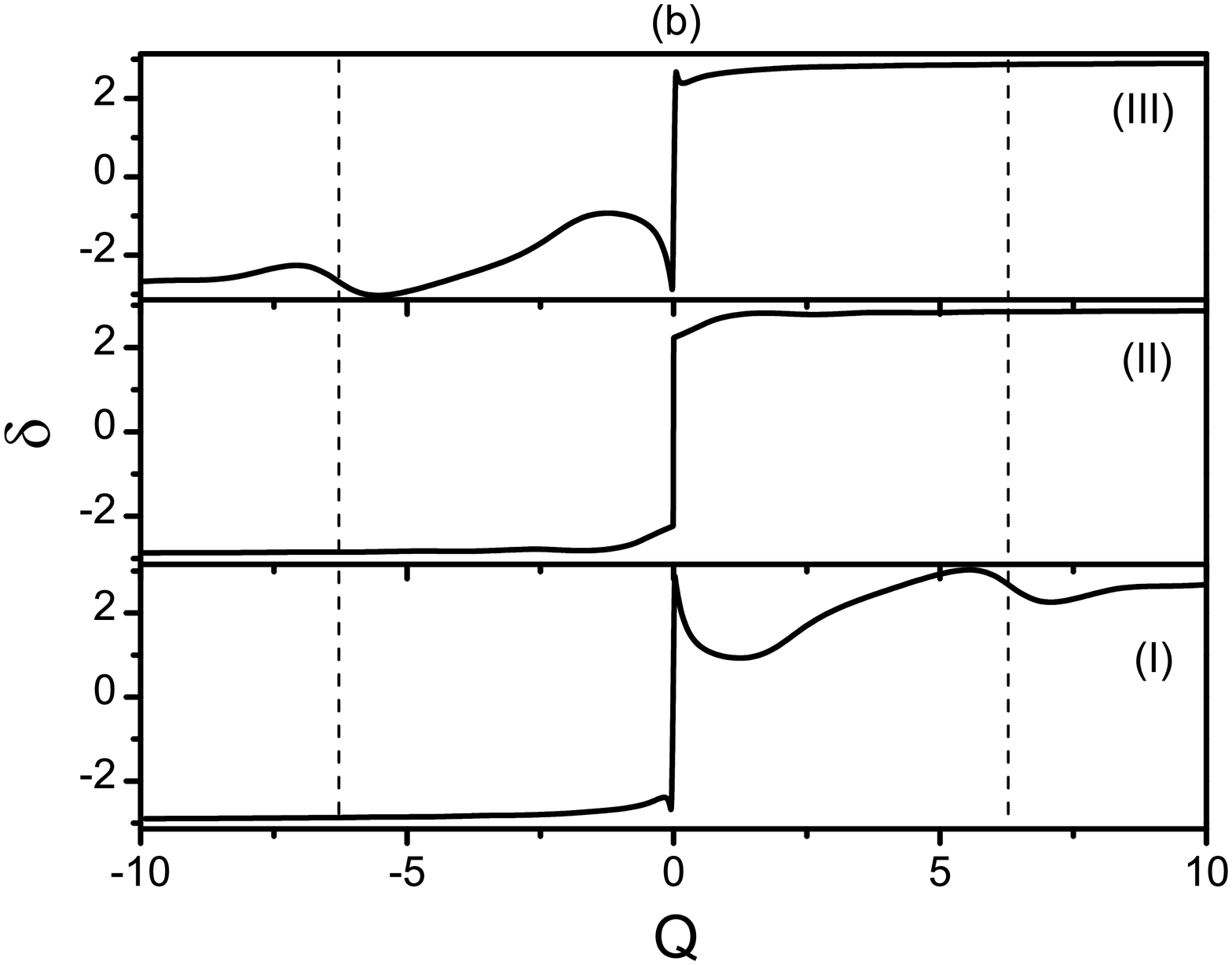}
\caption{  Relative amplitudes and phase difference of
the concentrations  (see Eq.~(\ref{N0-N1}) as functions of $Q$
for three spectrum branches. Panels I, II, III, are for $k= 1,0, -1$.
In (a), the full and dashed lines correspond to $\tilde{N}_{1}$ and $\tilde{N}_{0}$, respectively. }
\label{N0-N1-detales}
\end{figure}

In Fig.~\ref{N0-N1-detales} the relative amplitudes of concentrations of "isotropic",
and "anisotropic" electrons, and their phase difference $\delta (Q,k) \equiv \delta_0 (Q,k) -
\delta_1 (Q,k)$ are shown for three excitation branches presented in
Fig.~\ref{Omega-Q-with charge-1}.
The results illustrate useful symmetry properties of these functions.
Indeed, for the acoustic branch ($k=0$), the relative amplitudes,
$\tilde{N}_{0,1}(Q,k)/\tilde N (Q,k)$,
are even functions of $Q$, while the phase difference is an odd function of $Q$.
For a given optical branch, the relative amplitude is a strongly asymmetrical function
of $Q$. However, there are simple relationships between the characteristics
of two branches with $\pm k$. At given $k$ and $Q$, the relative amplitudes
coincides with those of the branch $-k$ at the wavevector $-Q$, while
$ \delta (Q,k) = -  \delta (-Q,-k)$. Significantly the different character of the excitations
corresponding to the acoustic and optical branches can be seen from
Fig.~\ref{N0-N1-detales} for small wavevectors. Indeed, for the acoustic branch
all functions, ${\tilde N}_0 (Q,0)\,,{\tilde N}_1 (Q,0)$ and ${\tilde N}  (Q,0)$ are of the same
order of magnitude while the phase shift $\delta \neq \pm \pi$. For the optical branches,
we obtain ${\tilde N}_0 (Q, k) \approx {\tilde N}_1 (Q, k) \gg \tilde{N} (Q, k)$ and $\delta \approx \pi$, i.e.,
the concentrations of the isotropic and streaming electrons oscillate in antiphase,
$\tilde N$ and the space charge are almost zero. For finite $Q$, the excitations are always
accompanied by a space charge wave. At $k = 1$ (-1) and $- 2 \pi < Q < 0$ $(0 < Q < 2 \pi)$,
the phase velocity of the excitation is negative and all disturbed electrons concentrations
are of the same order of magnitude. Otherwise, the phase velocity of the
excitation is positive.  For the developed streaming effect $(F_0 \gg F_{cr}), \,{ \tilde{ N}}_1 $
dominates over ${\tilde{ N}}_0$.

Thus, the accounting of space charge effects does not change considerably the basic properties
of the electron excitations under streaming transport regime. The multi-branch character of
the excitation spectra, the occurrence of the standing waves and other properties can be observed at
finite electron concentrations. The space charge accompanying the wave excitations affects
 mainly their damping. Space charge induced increase of damping of electron excitations
 is characteristic for regimes of electron motion with considerable dissipation.~\cite{remark-2}

\section{Discussion and Summary}

The cyclic electron motion caused by strongly inelastic processes (optical phonon emission)
is characteristic for many polar materials
and associated heterostructures subjected to high electric field
at low temperatures. The characteristic time period, $\tau_F$, and spatial period, $l_F$,
of this motion depend on the applied field, $F$.
In steady state regime, such a motion
reveals itself via the formation of a sufficiently anisotropic distribution
function and quasi-saturation of the drift velocity, the average energy, etc.
In time (frequency) domain, the cyclic electron motion gives rise
to the specific optical phonon transient time resonance at
frequencies defined by the time-of-flight, $\omega \approx 2 \pi/\tau_F$.
In real space, this kind of motion causes a spatial modulation of the electron
concentration and velocity with the period $l_F$, which has been
observed in structures of finite size.  It is believed that the cyclic
electron dynamics should be more pronounced in low-dimensional structures.

Using the BTE, we have developed an approximate approach
that
permits us to analyze {\em coupled spatio-temporal dynamics} of
two-dimensional electron gas under conditions of the cyclic motion.
As a results, we have found a novel type of  excitations of the drifting electron
gas.  These wave-like excitations are periodic in time and weakly damped
oscillations of the electrons in both real and momentum spaces.
Their frequency-wavevector relations consist of an infinite number
of continuous branches $\omega^k (q)\,,\,k=0, \pm 1, \pm 2...$.

The specific character of the correlated motion of electrons in the wave excitations strongly
depends on the branch number $k$ and the wavevector $q$. Thus, for $k=0\,, q  \rightarrow 0$, we
have obtained $\omega^0 \propto q$, the electron motion is phased, mainly,
in the real space, the excitation is conveyed  by a space charge wave. This excitation
can be defined as of acoustic type. For $k \neq 0$ and $q  \rightarrow 0$,
the excitations represent, mainly, strong oscillations in the momentum space
(time-dependent redistribution between 'isotropic' and 'streaming' groups of the electrons).
While the space charge is practically absent. These excitations can be defined
as of optical type. Each of the $\omega^k (q)$-branches once crosses the line
$\omega=0$ at $q = 2 \pi k/l_F$. This implies the existence of standing waves with
wavelength, $\lambda$, related to the spatial period of the cyclic motion as
$\lambda = l_F/k$. For the standing waves, electron motion in real and momentum spaces
is phased, but time independent. The found excitations are weakly damped.
Their damping is caused by  quasi-elastic scattering of electrons
and formation of a space charge.  The standing waves are undamped.

Now we shall discuss  the cyclic electron motion and the existence of the wave-like
excitations in real polar heterostructures with two-dimensional electrons.
For polar materials, the typical optical phonon energy is $30...100\,meV$. Due to the
exponential temperature dependence of the optical phonon absorption time
($\propto e^{-\hbar \omega_0/k_BT}$), this process is negligible at low
temperatures. For two examples given below, the absorption time is more
than $10^4$ times larger than the emission time  at $T \leq 100\,K$.
Other necessary and sufficient conditions for the realization of these effects are
 presented by Eqs.~(\ref{necessary-cond}) and (\ref{F-conditions}). These conditions
contain two material parameters,  the elastic scattering time in the
passive region, $\tau_p$, and the optical phonon emission  time, $\tau^{em}_{op}$.
At low temperatures, the former time can be estimated by the use of data on the low-field
mobility, $\mu_0$.
The latter time can be found, for example, exploiting the dielectric continuum model
of the optical phonons and the Frohlich electron-phonon interaction.~\cite{Ridley,Mitin}
To avoid effects of electron-electron collisions on the cyclic motion, the electron
concentration should be less than $10^{12}\,cm^{-2}$.~\cite{2DEG-OPTTR-2}

\begin{figure}
\includegraphics[height=9cm,width=9cm]{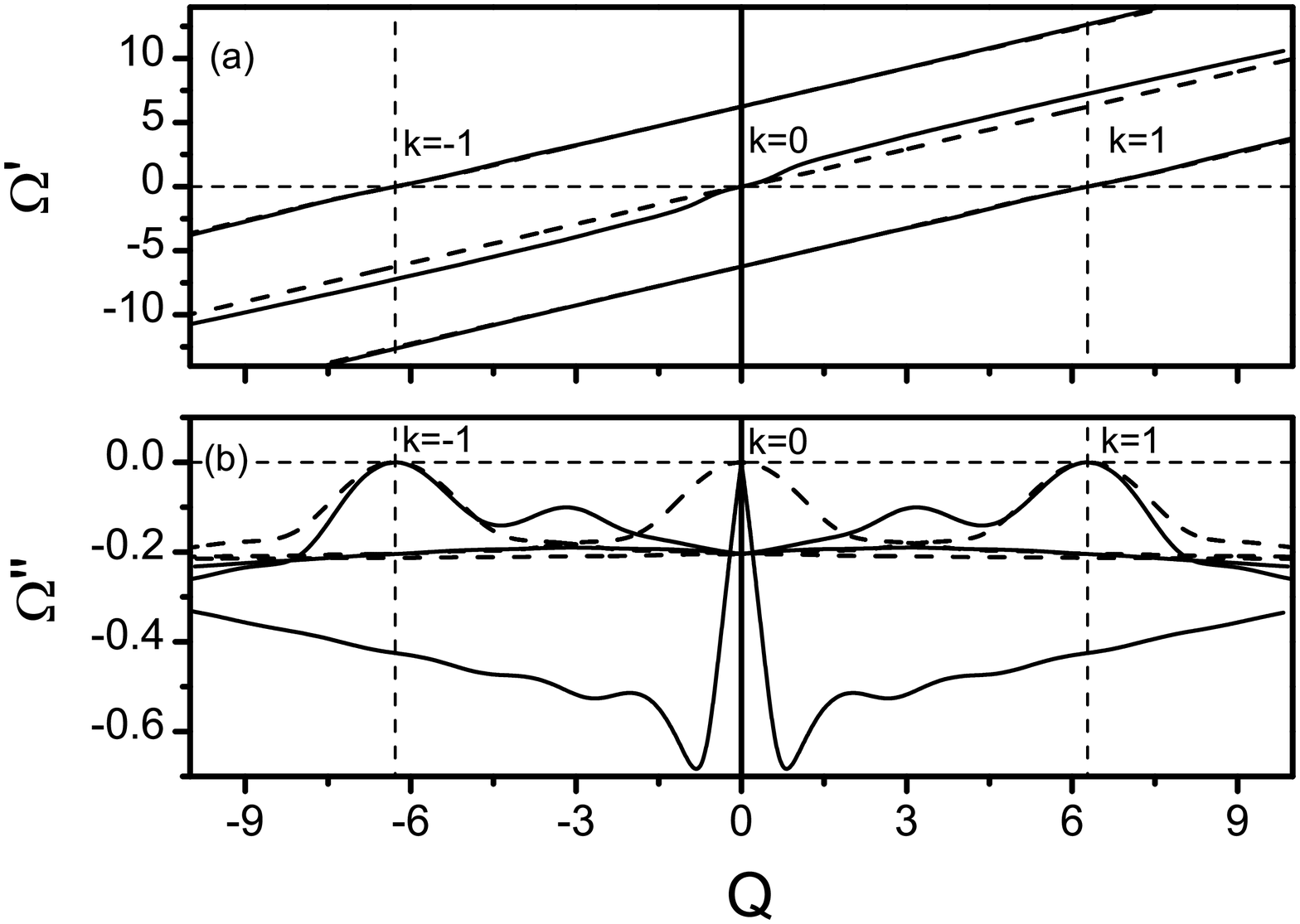}
\caption{The same as in Fig.~(\ref{set-branches-1}) at ${\cal N} =7.4$.
For GaN quantum wells, this corresponds to the electron concentration
$N=10^{12}\, cm^{-2}$.  In both panels,  for comparison the results at ${\cal N}=0$
are presented by the dashed lines. }
\label{Omega-Q-with charge-1}
\end{figure}

\begin{figure}
\includegraphics[height=9cm,width=9cm]{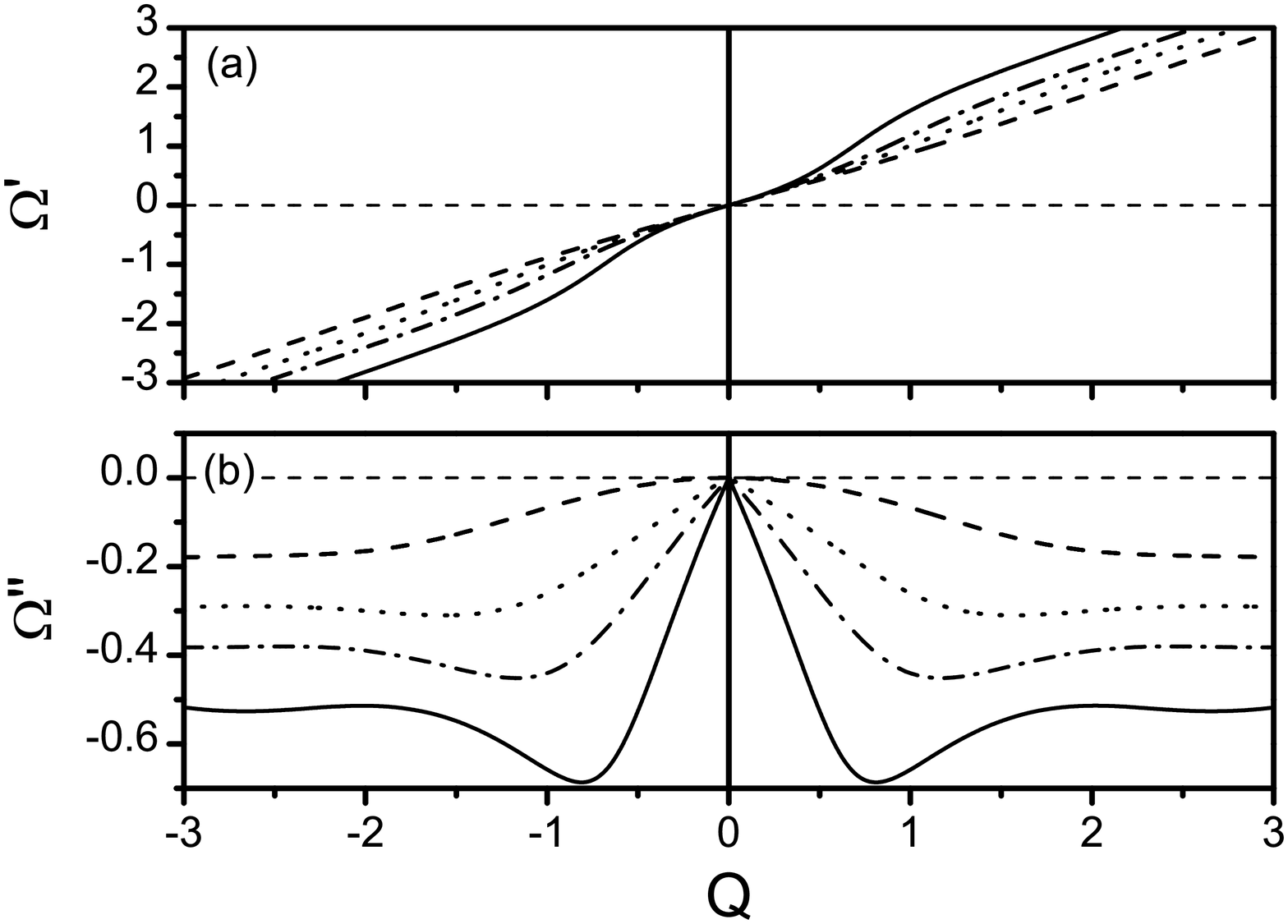}
\caption{ The same as in Figs.~\ref{set-branches-1} and \ref{Omega-Q-with charge-1}
for the acoustic spectrum  branch, $k=0$, at
different ${\cal N}$. For GaN quantum wells,  full, dash-dotted, dotted and dashed lines
correspond to  $N=10^{12},\,5 \times 10^{11},\, 2.5\times 10^{11} $ and $0$ (in $cm^{-2}$).}
\label{main-branch}
\end{figure}

First, we consider AlGaN/GaN heterostructures. For two-dimensional electrons in
GaN-channels, the optical phonon emission is estimated to be
$\tau^{em}_{op} \approx 0,01..0.02\, ps$.~\cite{2DEG-OPTTR,2DEG-OPTTR-3}
The highest low-field mobility was measured at sub-Helium temperatures for
structures with low density of dislocations:
$\mu_0 = (1...2) \times 10^5 cm^2/V s$.~\cite{Manfra-1, Porowski, Manfra-2}
These results were obtained for sufficiently low electron concentrations:
$N \approx (4...10) \times 10^{11}\,cm^{-2}$.  The measurements evidence
 that in such structures at moderately low temperature
the mobility is limited by acoustic phonon scattering.
At T = 50..100 K, the acoustic phonon limited mobility is
$\approx (5...2.5) \times 10^4 cm^2/V s$.~\cite{Porowski, Manfra-2, Faraon}
For numerical estimations, we choose a modest value of the mobility, $\mu_0 = 10^4  cm^2/V s$,
then we find $\tau_p \approx 1 ps$. Thus the strong inequality (\ref{necessary-cond}) is met.
For further calculations we use the following parameters:
$m^*=0.2 m_0$ ($m_0$ is the free electron mass), $\hbar \omega_0 = 92 \,meV\,,\,\kappa_0 =9$.
Then, according to Eq.~(\ref{cr-field}), the electric field for the
onset of the studied effects is $F_{cr} \approx 4.7\,kV/cm$. Thus, the numerical results shown in
Figs.~\ref{Omega-Q-with charge-1} and  \ref{main-branch} correspond to an applied field $F \approx 13.7\,kV/cm$.
The temporal and spatial periods of the cyclic motion are $\tau_F \approx 0.33\,\times 10^{-12}\,s$
and $l_F \approx 6.6 \times 10^{- 6}\,cm$, respectively. According to Eq.~(\ref{L}),
the characteristic electron concentration, for which the parameter ${\cal N} =1$,
is estimated to be $N_{ch} =1.4 \times 10^{11}\,cm^{-2}$. Having the scaling parameters $\tau_F,
\,l_f,\,N_{ch}$ and using the results of Figs.~\ref{Omega-Q-with charge-1}, \ref{main-branch} one
can recover the dimensional characteristics of the wave excitations. For example,
the frequency of the lowest optical-like branches at $q \rightarrow 0$ is found to be
$\omega^{\pm 1} (q \rightarrow 0) = 1.5 \times 10^{13} s^{-1}$, its damping equals $\gamma =
6 \times 10^{11} s^{-1}$. Undamped standing waves are realized for a wavevector
equal to $q = \pm 9.4 \times 10^{5} cm^{-1}$. For the same $q$, the frequency (the damping)
of the acoustic  and optical branches are $\omega \approx 2 \times 10^{13} s^{-1}$
($\gamma  \approx1.2 \times 10^{12} s^{-1}$) and  $\omega \approx 3.6 \times 10^{13} s^{-1}$
($\gamma \approx 6 \times 10^{11} s^{-1}$), respectively.

As second example of two-dimensional systems, where the discussed effects can be observed,
we consider ZnO/MgZn0 heterostructures.  For these strongly polar oxide heterostructures,
low field mobilities as high as $4 \times 10^5\,cm^2/V \,s$ were measured~\cite{ZnO-1}  at
sub-Helium temperature and electron concentrations $ (1...4) \times 10^{11}\,cm^{-2}$.
Above Helium temperature, it was found a phonon limited mobility temperature dependence:
$\mu_0 \propto T^{-3/2}$. Particularly, for $T \approx 10\, K$, it is achieved $\mu_0 \approx (2...4) \times
10^4\,cm^2/V\,s$.~\cite{ZnO-1,ZnO-2} For numerical estimates, we set $\mu_0 = 10^4  cm^2/V s\,,\,m^* = 0.3 m_0\,,\,
\hbar \omega_0 = 72\,meV\,,\,\kappa_0 = 8.1\,,\,\kappa_{\infty} = 3.7$.  Then, we obtain $\tau^{em}_{op} = 0.005...0.01\,ps$,
$\tau_p \approx 2 \,ps$, i.e., the necessary condition of Eq.~(\ref{necessary-cond}) holds. The critical field is
$F_{cr} = 2.9\,kV/cm$. For these parameters of the ZnO/MgZn0 heterostructure, results presented in
Figs.~\ref{Omega-Q-with charge-1} and  \ref{main-branch}  correspond to
 the applied field $F \approx 9.7\,kV/cm$. The periods of the relevant cyclic motion are
$\tau_F \approx 0.6\,\times 10^{-12}\,s$ and $l_F \approx 0.8 \times 10^{-5}\,cm$.
The characteristic electron concentration $N_{ch}$  equals
$ \approx 0.7 \times 10^{11}\,cm^{-2}$. With these scaling parameters and the results of Figs.~\ref{Omega-Q-with charge-1}, \ref{main-branch} one
can recover the dimensionless characteristics of the wave excitations.

Summarizing, we have analyzed the low-temperature behavior of nonequilibrium
two-dimensional electron gas in polar heterostructures subjected to moderately high
electric fields. At low temperatures, when the optical phonon emission is the fastest
relaxation process and an almost cyclic motion of individual electrons occurs, we have
found the existence of  collective wave-like excitations of the electrons.
These wave-like excitations are periodic in time and weakly damped
oscillations of the electrons in both real and momentum spaces.
The excitation spectra are of multi-branch character, each of the spectra branches
showing considerable  spatial dispersion. There are one acoustic-type and a
number of optical-type excitation spectra branches.
Their small damping is caused by  quasi-elastic
scattering of electrons and formation of a relevant space charge.
There are also waves with zero frequency and finite spatial periods - the standing
waves. The studied excitations of the electron gas can be also interpreted as
synchronous in time and real space manifestation
of the optical phonon transit time resonance.
Remarkably, these wave-like excitations exist in the electron gas even in the presence of significant
dissipation through optical phonon emission.
Estimations of the parameters of the excitations for two
examples of polar two-dimensional heterostructures, GaN/AlGaN and
ZnO/MgZnO, have shown that the excitation frequencies and the standing wave periods
are in THz-frequency range and sub-micrometer region, respectively.

\clearpage

\clearpage
\begin{references}

\bibitem{Shockley}
W. Shockley, Bell Syst. Tech. J. Eng. {\bf R 30},
990 (1951).

\bibitem{InSb}
Y. Katayama and K. F. Komatsubara, Phys. Rev. Lett. {\bf 19},
1421 (1967).


\bibitem{InGaAs}
P. F. Lu, D. C. Tsui, and H. M. Cox, Phys. Rev. Lett.
 {\bf 54}, 1563 (1985).

\bibitem{GaAs}T. W. Hickmott, P. M. Solomon, F. F. Fang, F. Stern, R. Fischer, and H.
Morkoc, Phys. Rev. Lett. {\bf 52}, 2053 (1984).
L. Eaves, P. S. S. Guimaranes, B. R. Snell, D. C. Taylor and
K. E. Singer, Phys. Rev. Lett. {\bf 55}, 262 (1985).

\bibitem{InP}
P. F. Lu, D. C. Tsui, and H. M. Cox, Phys. Rev.
 {\bf B 35}, 9659 (1987).


\bibitem{Reggiani-1}
V. Gruzinskis, P. Shiktorov, E. Starikov, L. Reggiani, L. Varani, and J. C.
Vaissiere, Semicond. Sci. Technol. {\bf 19}, S173 (2004).


\bibitem{Gonzalez-1}
A. I{\~ n}iguez-de-la-Torre, J. Mateos, and T. Gonzalez
J. Appl. Phys. {\bf 107}, 053707 (2010).

\bibitem{Andronov}
Gornik E and Andronov A A (ed) 1991 Opt. Quantum Electron.
23 S111–360 (Special Issue on Far-infrared Semiconductor
Lasers).

\bibitem{Reggiani-review}
E. Starikov, P. Shiktorov, V. Gruzinskis, L. Varani, et al.
J. Nanoelectron. Optoelectron {\bf 2}, 11 (2007).
E. Starikov, P. Shiktorov, V. Gruzinskis, L. Varani,
C. Palermo, J. F. Millithaler, and L. Reggiani,
J. Phys.: Condens. Matter {\bf 20}, 384209 (2008).


\bibitem{Vorobiev}
L. E. Vorob'ev, S. N. Danilov, V. N. Tulupenko and
D. F. Firsov, JETP Lett. {\bf 73}, 219 (2001).

\bibitem{2DEG-OPTTR}
 V. V. Korotyeyev, V. A. Kochelap, K. W. Kim and
D. L. Woolard, Appl. Phys. Lett. {\bf 82}, 2643 (2003).
K. W. Kim, V. V. Korotyeyev, V. A. Kochelap, A. A.Klimov  and
D. L. Woolard, J. Appl. Phys. {\bf 96}, 6488 (2004).

\bibitem{2DEG-OPTTR-2}
J. T. Lu, J. C. Cao and S. L. Feng, Phys. Rev. {\bf B 73}, 195326
(2006).
J. T. Lu and J. C. Cao, Semicond. Sci. Technol. {\bf 20}, 829 (2005).



\bibitem{2DEG-OPTTR-3}
E. Starikov, P. Shiktorov, V. Gruzinskis, A. Dubinov,
V. Aleshkin, L. Varani, C. Palermo, L. Reggiani,
J Comput Electron (2007) {\bf 6}, 45 (2007).
P. Shiktorov, E. Starikov, V. Gruzinskis, L. Varani, C. Palermo,
J-F. Millithaler and L. Reggiani, Phys. Rev. {\bf B 76}, 045333
(2007).

\bibitem{streaming-CNT}
A. Akturk, N. Goldsman, G. Pennington and A. Wickenden,
Phys. Rew. Lett. {\bf 98}, 166803 (2007);
A. Akturk, N. Goldsman and G. Pennington,
J. Appl. Phys. {\bf 102}, 073720 (2007);
A. Akturk, G. Pennington, N. Goldsman, and A. Wickenden,
IEEE Trans. Nanjtech. {\bf 6}, 469 (2007).


\bibitem{Baraff}
G. A. Baraff, Phys. Rev. {\bf 128}, 2507 (962).

\bibitem{Levinson}
I. I. Vosilius and I. B. Levinson, Sov. Phys. JETP {\bf 23}, 1104
(1966); {\bf 25}, 672 (1967).

\bibitem{Gribnikov}
Z. S. Gribnikov, V. A. Kochelap, Sov. Phys. JETP {\bf 31}, 562 (1970).

\bibitem{Kummer-function}
P. M. Morse,  and Feshbach, H. Methods of Theoretical Physics, Part I. New York: McGraw-Hill,  1953.


\bibitem{remark-1}
Analyzing Eq.~(\ref{Kampen1}) we mentioned that there is another kind of solutions
corresponding to "overdamped" excitations ($\Omega^{\prime} \sim \Omega^{\prime\prime}$).
For them we found also a set of branches, for which $d \Omega^{\prime}/d Q <0$, i.e. their group
velocities are negative.

\bibitem{remark-2}
Depending on frequency range, the space charge may lead to both additional
oscillatory phenomena (plasma oscillations at high frequency collisionless electron
motion) and additional relaxation (dielectric relaxation at low frequency dissipative
electron motion).

\bibitem{Ridley}
K. Ridley, Quantum Processes in Semiconductors (Clarendon,
Oxford, 1999).

\bibitem{Mitin}
V. V. Mitin, V. A. Kochelap, and M. A. Stroscio, Quantum Heterostructures
(Cambridge University Press, New York, 1999).

\bibitem{Manfra-1}
M. J. Manfra, K. W. Baldwin, A. M. Sergent, K. W. West, R. J. Molnar and J. Caissie
Appl. Phys. Lett., {\bf 85}, 5394 (2004); {\em ibid.},  {\bf 85},
1723 (2004); {\em ibid.}, {\bf 85}, 5279 (2004).


\bibitem{Porowski}
C. Skierbiszewski,  Z. Wasilewski, M. Siekacz, A. Feduniewicz, B. Pastuszka,
I. Grzegory, M.Leszczynski, and S. Porowski, Phys. Stat. Sol. (a) {\bf  201}, 320 (2004);
C. Skierbiszewski, K. Dybko, W. Knap, M. Siekacz, W. Krupczynski, G. Nowak, M. Bo.kowski,
J. Lusakowski, Z. R. Wasilewski, D. Maude, T. Suski and S. Porowski,
Appl. Phys. Lett., {\bf 86}, 102106 (2005).


\bibitem{Manfra-2}
E. A. Henriksen, S. Syed, Y. Ahmadian,
M. J. Manfra, K. W. Baldwin, and A. M. Sergent,
R. J. Molnar and H. L. Stormer, Appl. Phys. Lett. {\bf 86}, 252108 (2005)


\bibitem{Faraon}
A. Asgari, S. Babanejad and L. Faraone, J. Appl. Phis., {\bf 110}, 113713 (2011).


\bibitem{ZnO-1}
J. Falson, D. Maryenko, Y. Kozuka, A. Tsukazaki, and M. Kawasaki,
Appl. Phys. Exp. {\bf 4}  091101 (2011).


\bibitem{ZnO-2}
D.  G. Schlom and L. N. Pfeiffer, Nature Mater., {\bf 9} 881 (2010).
A. Tsukazaki, S. Akasaka, K. Nakahara, Y. Ohno, H. Ohno, D. Maryenko, A. Ohtomo
and M. Kawasaki, ibid, {\bf 9}, 889 (2010).



\end{references}
\end{document}